\pdfoutput=1

\documentclass[twocolumn,reprint,
nofootinbib,floatfix,preprintnumbers,citeautoscript,prd,superscriptaddress,amsmath,amssymb]{revtex4-1}

\usepackage{amsmath}
\usepackage{amssymb}
\usepackage[utf8]{inputenc}
\usepackage{graphicx}
\usepackage{hyperref}
\usepackage{color}
\usepackage{ascmac}
\usepackage{fancybox}

\newcommand{\waku}[2]{\medskip\begin{screen}{\bf #1}: #2 \end{screen}\medskip}
\newcommand{\wakutwo}[2]{\medskip\begin{shadebox}{\bf #1}: #2 \end{shadebox}\medskip}

\begin{document}

\title{Is Trans-Planckian Censorship a Swampland Conjecture?}

\author{Ryo Saito}
\affiliation{Graduate School of Sciences and Technology for Innovation, Yamaguchi University, Yamaguchi 753-8512, Japan}
\affiliation{Kavli Institute for the Physics and Mathematics of the Universe (WPI), University of Tokyo Institutes for Advanced Study, University of Tokyo, Chiba 277-8583, Japan}

\author{Satoshi Shirai}
\affiliation{Kavli Institute for the Physics and Mathematics of the Universe (WPI), University of Tokyo Institutes for Advanced Study, University of Tokyo, Chiba 277-8583, Japan}

\author{Masahito Yamazaki}
\affiliation{Kavli Institute for the Physics and Mathematics of the Universe (WPI), University of Tokyo Institutes for Advanced Study,  University of Tokyo, Chiba 277-8583, Japan}

\date{November, 2019}

\preprint{IPMU-19-0170}

\begin{abstract}
During an accelerated expansion of the Universe, quantum fluctuations of sub-Planckian size can be stretched outside the horizon and be regarded effectively classical. Recently, it has been conjectured that such horizon-crossing of trans-Planckian modes never happens inside theories of quantum gravity  (the trans-Planckian censorship conjecture, TCC). We point out several conceptual problems of this conjecture, which is in itself formulated as a statement on the restriction of possible scenarios in a theory: by contrast a standard swampland conjecture is a restriction of possible theories in the landscape of the quantum gravity. We emphasize the concept of swampland universality, i.e. that a swampland conjecture constrains any possible scenario in a given effective field theory. In order to illustrate the problems clearly we introduce several versions of the conjecture, where TCC condition is imposed differently to scenarios realizable in a given theory. We point out that these different versions of the conjecture lead to observable differences: a TCC violation in another Universe can exclude a theory, and such reduction of the landscape restricts possible predictions in our Universe. Our analysis raises the question of whether or not the trans-Planckian censorship conjecture can be regarded as a swampland conjecture concerning the existence of UV completion.
\end{abstract}

\maketitle



\section{Introduction}

The standard dogma of low-energy effective field theory and renormalization 
is based on the idea of separation of energy scales in physics. This has been a mixed blessing for physicists.
On the one hand such a hierarchy in energy scales ensures that, whenever one is interested in a low-energy effective field theory,
one can concentrate on the physics at the relevant energy scale: any ignorance concerning higher energy scales (such as whatever happens at the cutoff scale) 
can be parametrized by the coefficients of non-renormalizable operators in the effective theory.
On the other hand, the flip side of the same coin is that  it is in general
difficult to probe physics at very high energy scales, since these effects are typically 
suppressed by some powers of the cutoff scale. This challenge is 
extremely sharp for a quantum-gravity theorist who wishes to 
check their theories experimentally: what are the low-energy consequences of quantum gravity, whose energy scale can be as high as the Planck scale?

Early Universe provides an ideal setup for addressing this question.
It is widely believed in the literature that there is a period of accelerated expansion (inflation) in the early Universe \cite{Brout:1977ix,Starobinsky:1980te,Kazanas:1980tx,Guth:1980zm,Sato:1980yn}.
During inflation Planck-suppressed corrections to the inflaton potential
can spoil the flatness of the inflaton potential so that the physics can be very sensitive to Planck-scale physics.
The accelerated expansion in addition
stretches the sub-Planckian modes to
macroscopic scales, thus potentially opening up avenues for 
direct observations of Planck-scale physics. 
If this is indeed the case, one might argue that 
the cosmological fluctuations such as the Cosmic Microwave Background (CMB) anisotropies are highly sensitive to 
unknown physics at the Planck scale.
This is the famous trans-Planckian problem in inflation,
which has long been discussed in the literature,
see, e.g.\ Refs.~\cite{Martin:2000xs,Niemeyer:2000eh,Easther:2001fi,Kempf:2001fa,Bozza:2003pr}.
The trans-Planckian problem has also been long discussed for black holes, 
see e.g.\ Refs.~\cite{Unruh:1976db,Jacobson:1991gr,Jacobson:1993hn,Unruh:1994je,Polchinski:1995ta} for early references and Ref.~\cite{Jacobson:1999zk} for a summary.

Recently, Bedroya and Vafa boldly proposed that no cosmological trans-Planckian problem
arises in the landscape of the quantum gravity \cite{Bedroya:2019snp}---the horizon crossing of sub-Planckian modes
never happens. This conjecture is called the trans-Planckian censorship conjecture (TCC). 
In Ref.~\cite{Bedroya:2019snp}, TCC is claimed to be yet another example of a swampland conjecture \cite{Vafa:2005ui,Ooguri:2006in},
a necessary condition for the low-energy effective field theory to have  
a UV completion inside theories of quantum gravity, such as string theory (see e.g.\ Refs.~\cite{Brennan:2017rbf,Palti:2019pca,Yamazaki:2019ahj} for reviews on swampland conjectures).

After the proposal by Ref.~\cite{Bedroya:2019snp}, a number of papers \cite{Bedroya:2019tba,Cai:2019hge,Tenkanen:2019wsd,Das:2019hto,Mizuno:2019bxy,Brahma:2019unn,Draper:2019utz,Dhuria:2019oyf,Kamali:2019xnt,Torabian:2019zms,Cai:2019igo,Schmitz:2019uti,Kadota:2019dol,Berera:2019zdd,Brahma:2019vpl,Vafa:2019evj,Goswami:2019ehb,Okada:2019yne,Bernardo:2019bkz,Laliberte:2019sqc,Dasgupta:2019vjn,Lin:2019pmj,Brandenberger:2019jbs,Li:2019ipk,Seo:2019wsh,Kehagias:2019iem} have studied consequence of TCC. 
It should nevertheless be pointed out that there are still conceptual problems in TCC in itself.
In this paper, we would like to draw one's attention to problems in formulating TCC as a swampland conjecture.
In order to address this question more precisely, we formulate several versions of TCC and discuss the (de)merits of each in turn. 

\section{Swampland Conjectures}Let us begin with general remarks on swampland conjectures.

A swampland conjecture is a conjectural necessary condition for embedding a low-energy effective field theory
into a theory of quantum gravity. A swampland condition should therefore be 
satisfied for {\it all} possible scenarios theoretically realizable in 
a given effective field theory---it does not matter if the scenario is 
unlikely, or if it is realized in our Universe,
as long as the scenario has a non-zero probability of realization.
For later reference,
let us state this as a principle: 

\wakutwo{Swampland Universality}{A constraint from a swampland condition applies to {\it any} possible realization inside a given low-energy effective field theory.}

An example of a swampland conjecture
is the swampland de Sitter conjecture (dSC) \cite{Obied:2018sgi} (see also Refs.~\cite{Sethi:2017phn,Danielsson:2018ztv}).  
This conjecture states that the  total scalar potential
$V$ of a low-energy effective field theory satisfies the inequality
\begin{align}
M_{\rm Pl} |\nabla V| \ge c \, V \;,
\label{eq.dSC}
\end{align}
where $c$ is an $O(1)$ constant and $M_{\rm Pl}\simeq 2\times 10^{18} \textrm{GeV}$ is the reduced Planck mass. Since this condition 
applies to the potential of the theory and not to any particular 
realization, this condition indeed satisfies the swampland universality.

After the proposal of Ref.~\cite{Obied:2018sgi}, 
it has subsequently been pointed out that dSC causes severe problems with the presence of the Higgs field \cite{Denef:2018etk,Murayama:2018lie,Hamaguchi:2018vtv},
axion fields \cite{Murayama:2018lie,Choi:2018rze} as well as cosmic inflation models. \footnote{The literature is too vast to be exhaustively summarized here. See e.g.\ Refs.~\cite{Achucarro:2018vey,Kehagias:2018uem,Matsui:2018bsy,Garg:2018reu,Ben-Dayan:2018mhe,Kinney:2018nny,Murayama:2018lie} for early references based on dSC, and e.g.\ Refs.~\cite{Fukuda:2018haz,Chiang:2018lqx,Lin:2018rnx,Park:2018fuj,Cheong:2018udx,Kinney:2018kew,Haque:2019prw,Mizuno:2019pcm,Wang:2019eym,Channuie:2019vsp} for examples of similar analysis based on the refined dSC of Ref.~\cite{Ooguri:2018wrx}.}
For these reasons, there have been attempts to 
refine dSC \cite{Dvali:2018fqu,Andriot:2018wzk,Garg:2018reu,Ooguri:2018wrx,Murayama:2018lie,Andriot:2018mav}.
While the  inequality \eqref{eq.dSC} is speculative at a general point in the 
configuration space, it is better motivated in the 
asymptotic regions of the moduli space \cite{Dine:1985he,Maldacena:2000mw,Steinhardt:2008nk,Ooguri:2018wrx,Hebecker:2018vxz}. For this reason, one possible approach for refining the conjecture is to try to 
formulate another swampland conjecture which 
relaxes the constraint from \eqref{eq.dSC} at a general point in the moduli space,
while preserving the essence of \eqref{eq.dSC} in the asymptotic regions. 
TCC can be regarded as an example for such an attempt.

\section{Trans-planckian Censorship Conjecture}Let us consider a quantum fluctuation of sub-Planckian size
(of length scale $l\lesssim l_{\rm Pl}$, where $l_{\rm Pl}$ is the Planck length).
The expansion of the Universe stretches the physical scale of this fluctuation by a factor $a/a_{\rm ini}$, where $a_{\rm ini}$ and $a$ denote the scale factor at the initial and given time respectively. 
When the expansion accelerates, 
this factor grows monotonically with respect to the Hubble horizon $\sim H^{-1}$.
Therefore, the physical scale will eventually exit the Hubble horizon if the accelerated expansion continues for a sufficiently long time. 
TCC claims that this horizon exit of the trans-Planckian quantum fluctuation never occurs \cite{Bedroya:2019snp}:
\begin{align}
l_{\rm Pl} \frac{a}{a_{\rm ini}} \lesssim \frac{1}{H} \,, \quad \textrm{i.e.\ } \quad
\frac{a}{a_{\rm ini}} \lesssim \frac{M_{\rm Pl}}{H}\;.
\label{eq.TCC}
\end{align}
We will hereafter refer to the inequality of Eq.~\eqref{eq.TCC} as the TCC condition.

When a quantum fluctuation exits the Hubble horizon, it is widely believed that 
the quantum state transits into an effectively classical state (see e.g.\ Refs.~\cite{Starobinsky:1986fx,Brandenberger:1992sr,Polarski:1995jg,Lesgourgues:1996jc,Kiefer:1998qe,Barvinsky:1998cq} for a sample of early references). 
The typical narrative is that the state is in a highly-squeezed state
after the horizon exit and then quickly decoheres due to interactions with the environment.
We do not need detailed mechanisms for this quantum-to-classical transition in the following, 
except we should keep in mind that
the point of TCC concerns this classicalization, rather than the horizon exit itself.

As already stated, TCC can be regarded as an upgrade of dSC. 
TCC is in general weaker than dSC at a generic point of the configuration space (thus avoiding the problems of dSC discussed above), while preserving the essence of dSC 
at asymptotic regions \cite{Bedroya:2019snp} (cf.\ \cite{Dine:1985he,Ooguri:2018wrx,Hebecker:2018vxz}). 

\begin{figure}[htbp]
    \centering
   \includegraphics[width=0.48\textwidth]{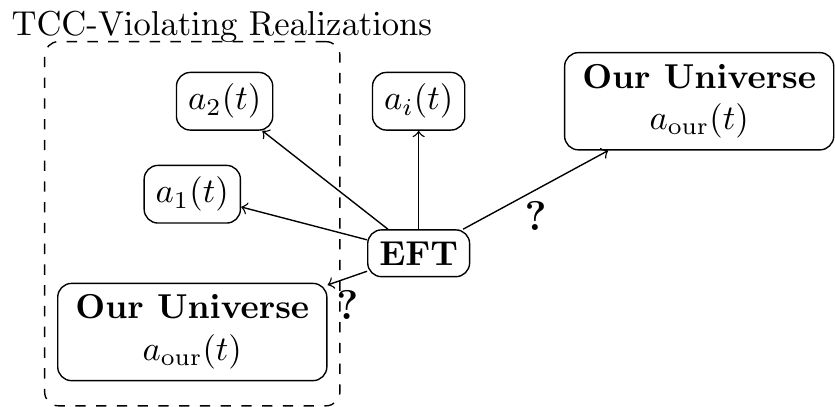}
    \caption{An example of EFT in the swampland when the swampland universality is applied for TCC. This theory is excluded because it allows realizations that do not satisfy the TCC condition \eqref{eq.TCC} (TCC-violating realizations), irrespective of whether our Universe is in them or not. Such reduction of the landscape will also restrict possible predictions in our Universe.}
    \label{fig:universality}
\end{figure}

\begin{figure}[htbp]
   \centering
   \includegraphics[width=0.46\textwidth]{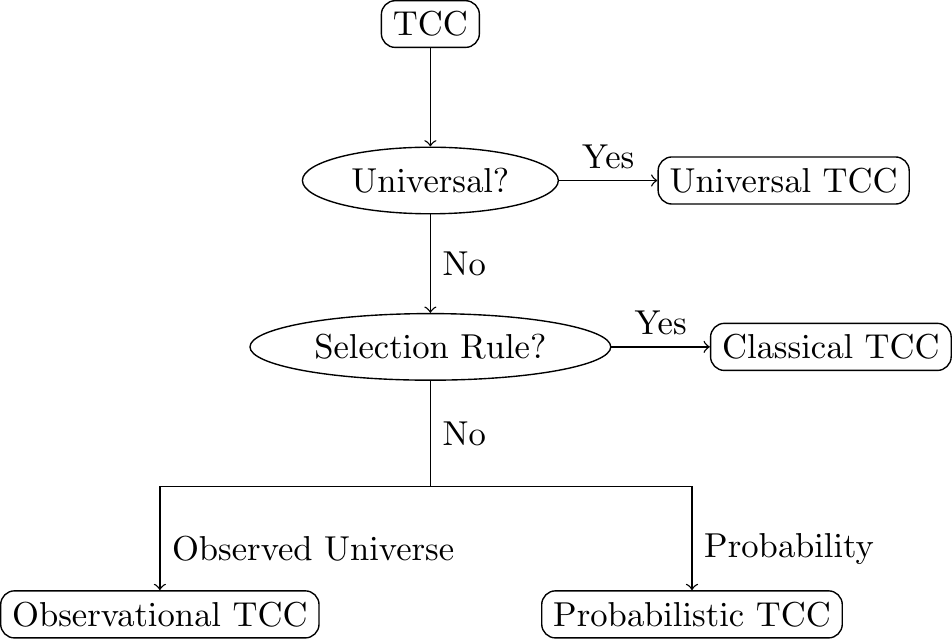}
    \caption{For precise discussions of TCC, it is important to 
    distinguish between different variations of TCC listed in this figure, since each leads to different conclusions and interpretations. 
    We can first ask if TCC applies universality, which is the case for the universal TCC (UTCC).
    Another possibility is to impose some selection rule and apply TCC to some subsector, e.g.\ by restricting to classical solutions.
    This gives the classical TCC (CTCC). Another option is to require that there is at least
    one TCC-consistent Universe realizable inside the theory (observational TCC, OTCC). This is minimally enough for the 
    assumption that TCC applies to our observable Universe, irrespective of what happens in other possible scenarios.
    Finally, we can adopt probabilistic interpretations, and allow for TCC violations with ``small' probabilities (probabilistic TCC, PTCC).}
    \label{fig:flowchart}
\end{figure}

Despite these similarities, there is actually a huge difference between TCC
and dSC. While dSC is a constraint on the potential of a {\it theory},
and hence of the theory itself, TCC is a constraint on a {\it realization} (a scenario) inside the theory, at least on a first look.
This is in some tension with the expectation that a swampland condition should be a constraint for the low-energy effective field theories.
A given low-energy effective field theory in general allows for many realizations, 
some of which might violate TCC even when others are consistent with TCC. 
The swampland universality implies that such an effective field theory is in the swampland, 
irrespective of whether our particular Universe satisfies the TCC condition \eqref{eq.TCC} (Fig.~\ref{fig:universality}). 
The possible existence of a TCC-violating Universe might seem like an academic problem and irrelevant to us.
It has, however, an observable impact: a stronger reduction of the quantum-gravity landscape leads to more restrictions on possible predictions in our Universe.
In order to see if this is indeed the case, 
we first need to be more explicit on what TCC means, 
since TCC can be interpreted in several different manners.
In the following we discuss each interpretation in turn (see Fig.~\ref{fig:flowchart}). 
We will in particular see that some versions put the Standard Model of particle physics and cosmology into the swampland.

\section{Universal TCC} \label{s:utcc}
One natural interpretation of TCC
is the following version, which straightforwardly satisfies the swampland universality:
\waku{Universal TCC (UTCC)}{the TCC condition \eqref{eq.TCC} is obeyed for {\it classically-observable quantities in any possible realization} allowed in a given effective field theory.}

Note that here we included the phrase ``classically-observable quantities'' in the formulation of UTCC,
since an intrinsically-quantum Universe does not have a well-defined concept of the classical spacetime
(and hence of the classical scale factor),  in which case it is not clear how the TCC condition \eqref{eq.TCC}
should be formulated.

\begin{figure}[htbp]
    \centering
   \includegraphics{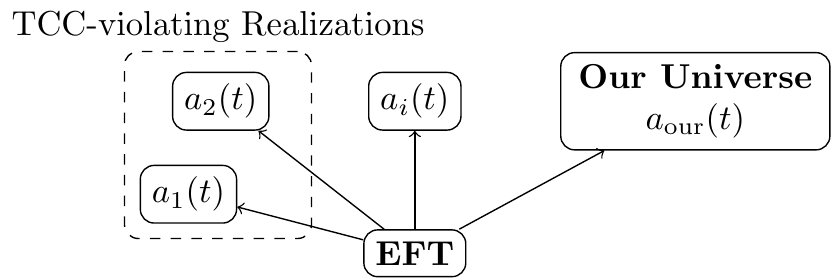}
    \caption{UTCC claims that the TCC condition \eqref{eq.TCC} is obeyed for classically-observable quantities in any possible realization allowed in a given effective field theory. This condition is violated if the theory admits even one TCC-violating Universe as a possibility.}
    \label{fig:UTCC}
\end{figure}

\begin{figure}[htbp]
    \centering
   \includegraphics[width=0.47\textwidth]{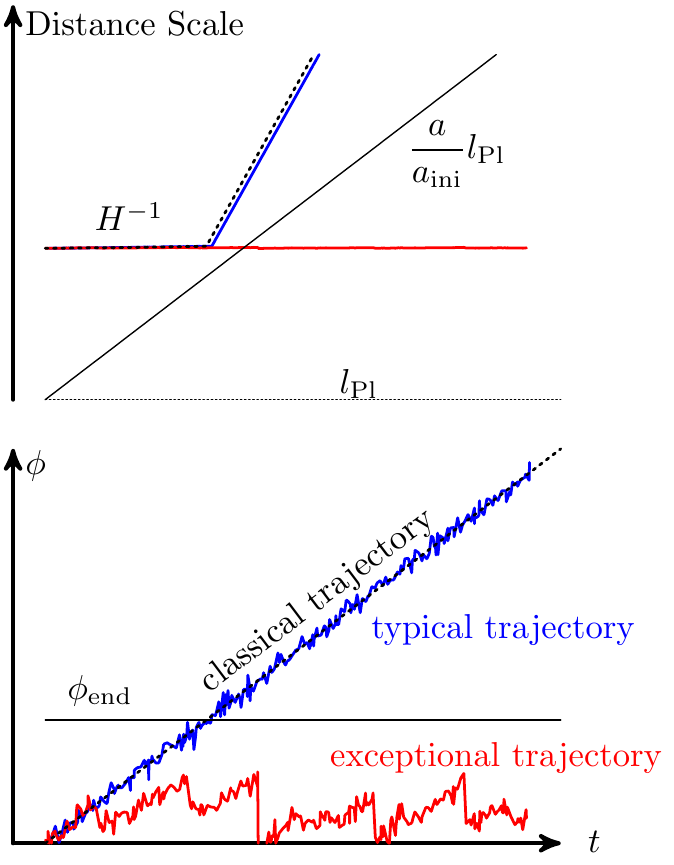}
    \caption{Schematic picture of stochastic trajectories of the inflaton field $\phi$ due to the backreaction effect. 
    The inflation ends when the field value crosses $\phi=\phi_{\rm end}$.
    The exceptional trajectory (red line) violates the TCC condition \eqref{eq.TCC}, while the typical trajectory (blue line) satisfies it.}    \label{fig:fluc}
\end{figure}

This version of TCC is rather strong and 
essentially excludes {\it any} inflationary scenario, 
while the condition \eqref{eq.TCC} by itself allows for a finite period of inflation. 
This is because, in this version of TCC, the TCC condition \eqref{eq.TCC} should be applied to {\it all} possible trajectories of the scale factor in a given theory---one can consider a TCC-violating realization in any inflationary model by taking into account the backreaction of the fluctuations to the background expansion. 
After the horizon exit, as mentioned above, 
the fluctuations are classicalized and locally indistinguishable from the corresponding background.
As known in the stochastic inflation formalism
\cite{Vilenkin:1983xq,Starobinsky:1986fx,Rey:1986zk,Goncharov:1987ir,Nambu:1987ef,Nakao:1988yi,Nambu:1988je,
Kandrup:1988sc,Nambu:1989uf,Salopek:1990re,Mollerach:1990zf,Mijic:1990qx,Linde:1993xx,Starobinsky:1994bd},
the backreaction of the fluctuations changes the background trajectory stochastically. 
It is then possible, for example, that the background inflaton stays at the same point in the potential for an arbitrary long time due to the backreaction effect, 
making the value of the e-folding number of the inflation arbitrary large (see Fig.~\ref{fig:fluc}). 
The TCC condition \eqref{eq.TCC} is clearly violated in this situation. 
Such a large backreaction effect is familiar in the eternal inflation scenario \cite{Linde:1982ur,Steinhardt:1982kg,Vilenkin:1983xq,Linde:1986fc,Linde:1986fd,Goncharov:1987ir}, 
except here we do not necessarily assume that the inflation continues in most part of our Universe.
In any inflationary model, a realization of large quantum fluctuations exists with a non-zero probability 
and their backreaction effect leads to the violation of the TCC condition \eqref{eq.TCC}. 
It is also easy to understand that any inflation model with a metastable vacuum is excluded by a similar argument. 
To elude this conclusion, an inflationary model should strictly forbid all TCC-violating realizations, 
while allowing realizations of a finite period of inflation with the e-folding number \eqref{eq.TCC} 
and the production of the density fluctuations in our Universe. 
To our knowledge, such inflationary model has not been known.
It should be remarked that this argument is applicable to all scalar fields with a positive and flat potential. 

Moreover, there is a more serious objection that UTCC is incompatible with {\it any} scalar field whose potential has a local maximum
with positive energy. 
This is because we can consider a realization where the scalar field is placed on the top of the potential and stays there forever. 
In this solution, the accelerated expansion occurs eternally, 
leading to the violation of the TCC condition \eqref{eq.TCC}.
This is indeed the case for the Higgs potential
in the Standard Model of particle physics. 
As in the case of the de Sitter swampland conjecture, 
one can try to rectify this problem
by e.g.\ modifying the electroweak sector by including extra fields 
or by coupling the Higgs field to the quintessence field \cite{Ratra:1987rm,Wetterich:1987fm,Zlatev:1998tr}. 
However, each has its phenomenological problems as already pointed out in Refs.~\cite{Denef:2018etk,Murayama:2018lie,Hamaguchi:2018vtv} (save for rather fine-tuned possibilities). The same applies to local maxima of mesons or axions \cite{Murayama:2018lie,Choi:2018rze}.

We therefore conclude that, while theoretically appealing as a swampland conjecture,
UTCC is in practice too strong to be imposed.

Now, we know that the swampland universality of TCC cannot be satisfied in its literal form. 
In the following, then, we will examine three possibilities to formulate TCC in an acceptable form: 
(1) TCC-violating solutions are not realized due to (unknown) quantum gravity effects, 
(2) TCC-violating solutions can exist but are never observed by us, 
(3) TCC violations with a small probability are tolerated.   

\section{Classical TCC}\label{s:ctcc}

The first approach is to introduce a selection rule: one assumes that only a limited class of realizations is actually allowed in a given theory. 
In order that this formulation of TCC qualifies as a swampland conjecture, this selection rule should be uniquely determined given an effective field theory. 
However, as we will see, 
it would be difficult to find a reasonable selection rule. 
To clarify this point, let us consider a particular selection rule that only picks up classical realizations:

\waku{Classical TCC (CTCC)}{the TCC condition \eqref{eq.TCC} is obeyed for {\it any possible classical realization (i.e. classical solution)} allowed in a given effective field theory.}

\begin{figure}[htbp]
    \centering
   \includegraphics{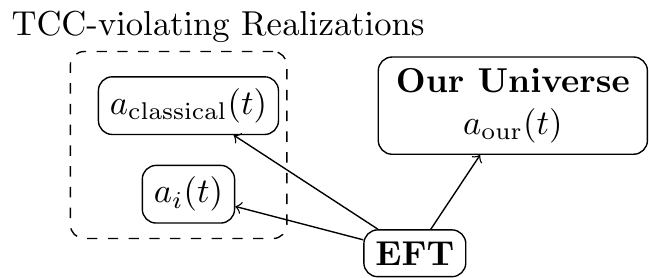}
    \caption{An example of EFT in the swampland for CTCC. CTCC claims that the TCC condition \eqref{eq.TCC} is obeyed for  any possible classical realization allowed in a given effective field theory. This condition is therefore violated only when the theory contains a TCC-violating classical realization, as shown in this figure.}
    \label{fig:CTCC}
\end{figure}

In this conjecture, only the classical realizations are assumed to be possible and the swampland universality is applied to them. 
CTCC gives a proper swampland condition because, for a given effective field theory, its possible classical solutions are unambiguously defined as the extrema of its action integral. 
It is obvious that CTCC is unacceptable without any proviso because quantum effects have been observed in experiments. 
A possible excuse is to assume that CTCC is only applied to cosmological solutions, 
where quantum gravity comes into play. 

However, there remain problems with this CTCC. 
First, CTCC has to at least forbid the inflaton's fluctuations that cause the backreaction. 
Therefore, CTCC forces us to abandon the standard inflationary paradigm of the structure formation from the quantum fluctuations. 
One might argue that the cosmological fluctuations can be seeded by other scalar fields such as the curvaton \cite{Linde:1996gt,Enqvist:2001zp,Lyth:2001nq,Moroi:2001ct}, 
but the swampland universality (even its classical version) implies that their fluctuations should be also forbidden 
because any such scalar field can be the inflaton---there is a classical solution where the field drives inflation. 

Another point is that CTCC misses the motivations of TCC. Recall that TCC is meant to 
preclude the classicalization of the trans-Planckian quantum fluctuations 
in order to preserve a self-consistent classical picture of spacetime on larger scales. 
On the other hand, CTCC forbids these trans-Planckian quantum fluctuations as well as the problematic fluctuations that cause the backreaction effect.  
Therefore, it is unclear what is problematic when CTCC is violated. 

Furthermore, the problem of the local maxima of the Higgs and meson potentials, as pointed out above for UTCC, still remain even in this formulation.

From the arguments above, the selection rule should strictly forbid large quantum fluctuations that stop a field rolling down its potential slope 
but should allow small quantum fluctuations that kick a field from a local maximum of the potential, 
and those that source the cosmological fluctuations for the structure formation if one wishes to keep the standard inflationary paradigm. 
If we brutally forbid large quantum fluctuations, 
it is very likely to be in sharp conflict with basic principles of the quantum mechanics---since the Schr\"odinger equation is diffusive, the wavefunction spreads over the field space unless confined by an infinite potential barrier. One can still be very speculative and argue that it might be possible to justify such a selection rule in the grand framework of quantum gravity---a possible analogy is that the quantization condition in the Bohr's old quantum theory, a selection rule of classical trajectories, was justified eventually in the modern quantum mechanics in a framework completely different from classical mechanics. 
While this could be possible in principle, one should note that this is a dramatically more radical requirement than that required by the standard swampland conjectures.

\section{Observational TCC} \label{s:otcc}

We have discussed difficulties in introducing a selection rule for realizations. 
Then, a next possible approach is to allow all realizations, but forbid that the TCC violation is observed, at least, by us: 

\waku{Observational TCC (OTCC)}{the TCC condition \eqref{eq.TCC} is obeyed in our Universe.}

\begin{figure}[htbp]
    \centering
   \includegraphics{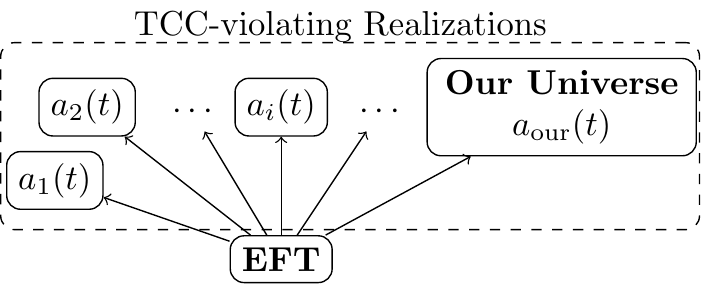}
    \caption{An example of EFT in the swampland for OTCC. OTCC claims that the TCC condition \eqref{eq.TCC} is obeyed in our Universe, 
    and an EFT is excluded 
    only when our Universe violates the TCC condition, irrespective of what happens in other possible realizations. 
    In other words, an EFT is allowed if it has at least one TCC-consistent solution as a possibility. The swampland universality is maximally violated for OTCC.}
    \label{fig:OTCC}
\end{figure}

This is the version implicitly used in many of the recent papers discussing phenomenological consequences of TCC. 
OTCC provides the minimum setup to discuss practical consequences of TCC. 
As we will see, however, some results from the literature are reproduced here but some are {\it not} in OTCC. 

Let us first discuss phenomenological predictions of OTCC. 
The first example is an upper limit on the primordial tensor amplitude, or the energy scale of inflation, 
as already discussed elsewhere \cite{Das:2019hto,Mizuno:2019bxy,Brahma:2019unn,Dhuria:2019oyf,Kamali:2019xnt,Torabian:2019zms,Kadota:2019dol,Berera:2019zdd,Goswami:2019ehb}. 
Applying the TCC condition \eqref{eq.TCC} to the end time of inflation, 
we find that the expansion rate at this time $H_{\rm e}$ is limited by the e-folding number of inflation $N_{\rm inf}$ as,
    \begin{align}
        \frac{H_{\rm e}}{M_{\rm Pl}} \lesssim e^{-N_{\rm inf}} \,.  \label{eq.TCC_inflation}
    \end{align}
When the expansion rate is approximately constant during inflation $H_{\rm e} \simeq e^{-\epsilon_H N_{\rm inf}} H_{\rm inf}~(\epsilon_H \ll 1)$, 
this immediately implies that the longer inflation leads to more suppression of the primordial tensor fluctuations:
    \begin{align}
        {\cal P}_T = \frac{2}{\pi^2} \left(\frac{H_{\rm inf}}{M_{\rm Pl}}\right)^2 < 0.2\, e^{-(2-\epsilon_H)N_{\rm inf}}\,, \label{eq:rupper}
    \end{align}
or equivalently the energy scale of inflation, $V/M_{\rm Pl}^4 \simeq (3\pi^2/2){\cal P}_T$.
On the other hand,  
the inflation should be long enough for the CMB modes to cross the horizon during inflation as, 
    \begin{align}
        e^{N_{\rm inf}} > \left(\frac{H_{\rm inf}}{H_0}\right) e^{-N_{\rm after}} \,, \label{eq:nlower}
    \end{align}
in order that the fluctuations are created from the quantum fluctuations.
Here, $H_0$ and $N_{\rm after}$ are the Hubble parameter at the present time and the e-folding number after the inflation, respectively. 
Therefore, since $N_{\rm after}$ cannot be arbitrarily large, 
it is possible to find an upper limit on the primordial tensor amplitude.

 The precise value of the upper limit depends on the history of Our Universe. 
 The e-folding number $N_{\rm after}$ is decomposed into those of the inflaton oscillation $N_{\rm osc}$ and of the Big Bang Universe,
     \begin{align}
        e^{N_{\rm BB}} &= \left(\frac{g_{*s}(T_R)}{g_{*s}(T_0)}\right)^{\frac{1}{3}} \left(\frac{T_R}{T_0}\right) \nonumber \\
        &= \Omega_{\rm rad}^{\frac{1}{4}}\left(\frac{g_{*s}(T_R)}{g_{*s}(T_0)}\right)^{\frac{1}{3}}\left(\frac{g_{*}(T_0)}{g_{*}(T_R)}\right)^{\frac{1}{4}} \left(\frac{H_R}{H_0}\right)^{\frac{1}{2}}\,,
    \end{align}
where the subscripts $R$ and $0$ indicate quantities at the reheating and the present time, respectively, and $g_{*s}(T)$ ($g_{*}(T)$) is the effective degrees of freedom for the entropy density (energy density) at temperature $T$.
When the instantaneous reheating is assumed, we have $N_{\rm osc} = 0$ and $H_R = H_e \simeq H_{\rm inf}$.
Then, from Eqs. \eqref{eq:rupper} and \eqref{eq:nlower}, one can find an extremely small upper limit on the tensor-to-scalar ratio \cite{Bedroya:2019tba},
    \begin{align}
        r \equiv \frac{{\cal P}_T}{{\cal P}_{\cal R}} < \frac{2}{\pi^2 {\cal P}_{\cal R}}\left(\frac{1}{\Omega_{\rm rad}}\frac{H_0^2}{M_{\rm Pl}^2}\right)^{\frac{1}{3}} = {\cal O}(10^{-30}) \,, \label{rupper_inst}
    \end{align}
or $V^{1/4} < {\cal O}(10^8){\rm GeV}$ on the energy scale of inflation, for the values of the cosmological parameters reported in Ref. \cite{Aghanim:2018eyx}. 
Here, we have omitted the effective degrees of freedom. 
This tight constraint can be relaxed by considering, e.g., multi-stage inflation $H_{\rm e} \neq H_{\rm inf}$ or non-standard history of the Universe $N_{\rm after} \neq N_{\rm BB}$ \cite{Das:2019hto,Mizuno:2019bxy,Brahma:2019unn,Dhuria:2019oyf,Kamali:2019xnt,Torabian:2019zms,Kadota:2019dol,Berera:2019zdd,Goswami:2019ehb}.
A general argument with them leads to \cite{Mizuno:2019bxy}, 
    \begin{align}
        r < \frac{g_{\ast}(T_0)}{45 \Omega_{\rm rad}{\cal P}_{\cal R}}\left(\frac{T_0}{T_R}\right)^2= {\cal O}(10^{-8})\left(\frac{T_R}{1~{\rm MeV}}\right)^{-2} \,. \label{rupper_general}
    \end{align}

A tighter constraint can be obtained when one seriously considers the flatness problem. 
Imposing the observed upper bound on the curvature density parameter $\Omega_{K0}<\Omega_{K,{\rm obs}}$, 
its initial value $\Omega_{K,{\rm ini}}$ should satisfy,
    \begin{align}
        e^{N_{\rm inf}} > \left|\frac{\Omega_{K,{\rm ini}}}{\Omega_{K,{\rm obs}}}\right|^{\frac{1}{2}}\left(\frac{H_{\rm inf}}{H_0}\right) e^{-N_{\rm after}} \,, \label{eq:curvature}
    \end{align}
which is obtained from Eq. \eqref{eq:nlower} with the replacements $H_{\rm inf} \to |\Omega_{K,{\rm ini}}|^{1/2}H_{\rm inf}$ and $H_0 \to |\Omega_{K,0}|^{1/2}H_0$.
From a similarity to Eq. \eqref{eq:nlower}, 
it is easy to find that 
    \begin{align}
        r < \left|\frac{\Omega_{K,{\rm obs}}}{\Omega_{K,{\rm ini}}}\right| r_{\rm upper} \,,
    \end{align}
where $r_{\rm upper}$ is the upper limit in Eq. \eqref{rupper_general}.
Therefore, unless one fine-tunes the value of $\Omega_{K,{\rm ini}}$, the present constraint $\Omega_{K0} < \Omega_{K,{\rm obs}} = {\cal O}(10^{-3})$ \cite{Aghanim:2018eyx} leads to a tighter constraint on $r$. 
We also conclude that the spatial curvature $\Omega_{K0}$ should be detectable in future \cite{Mao:2008ug} when the primordial tensor modes with $r \sim r_{\rm upper}$ are observed.
Note that a negative value of $\Omega_K$ is claimed to arise generically in the string theory landscape \cite{Freivogel:2005vv} (see also Ref. \cite{Guth:2012ww}),
when we consider a quantum tunneling as in the open inflation scenario
\cite{Bucher:1994gb,Yamamoto:1995sw,Linde:1995xm,Linde:1995rv}.
In this respect, it would be interesting to further study constraints on the spatial curvature. 

Here, it should be noted that these constraints are only applied to the primordial tensor modes generated by the standard mechanism. 
First, for the tensor-to-scalar ratio with such a small amplitude, 
it is necessary to take into account non-linearly generated tensor modes \cite {Cook:2011hg, Mollerach:2003nq, Baumann:2007zm, Martineau:2007dj} as well as foreground contamination. 
Moreover, 
several non-standard mechanisms are known to generate large tensor modes \cite{Barnaby:2012xt, Adshead:2013nka,Biagetti:2013kwa, Biagetti:2014asa, Dimastrogiovanni:2016fuu}. 
These non-standard mechanisms should be also tested with the swampland conjectures as well as observations of e.g.\ the scalar perturbations \cite{Ozsoy:2014sba, Fujita:2014oba, Mirbabayi:2014jqa, Fujita:2017jwq}.
A possible viable example is the generation from spectator non-Abelian (e.g.\ $\mathrm{SU}(2)$) gauge fields coupled to a pseudo scalar axion \cite{Adshead:2012kp, Adshead:2013nka}. 
In this mechanism, quantum fluctuations of the gauge fields can be enhanced through a transient instability and source large tensor modes. 
It has been argued in Ref. \cite{Fujita:2017jwq} that the tensor-to-scalar ratio can reach to $r \sim 10^{-3}$ at a specific scale even for a very low energy scale of inflation $V^{1/4} \simeq 30~{\rm MeV}$, which is well below the upper limit \eqref{rupper_inst}.  
This scenario should be further investigated if it is consistent with the other swampland conjectures and can be realized in the quantum-gravity landscape (see e.g.\ Ref.~\cite{Agrawal:2018mkd} for discussion).
It would be interesting to explore this point further.

The observational TCC also constrains the maximum value of the reheating temperature, $T_R$. 
Since the maximum reheating temperature is realized for the instantaneous reheating, 
the inequality \eqref{rupper_inst} can be rewritten as the upper limit on the reheating temperature \cite{Bedroya:2019tba},
    \begin{align}
        T_R < {\cal O}(10^8)~ {\rm GeV} \,,
    \end{align}
or
    \begin{align}
        T_R <{\cal O}(10^8 )  \left|\frac{\Omega_{K,{\rm obs}}}{\Omega_{K,{\rm ini}}}\right|^{\frac{2}{3}}  ~{\rm GeV}\,, \label{eq.Omega_K}
    \end{align}
when the flatness problem is taken into account. 
These are in tension with the thermal leptogenesis \cite{Fukugita:1986hr}, which requires $T_R\gtrsim 10^9 \textrm{GeV}$ \cite{Giudice:2003jh,Buchmuller:2004nz}.

Next, 
let us point out that OTCC does {\it not} reproduce all known predictions in the literature. 
To clarify this point, let us formulate OTCC as a swampland condition. 
For the TCC condition \eqref{eq.TCC} to be satisfied in our Universe, 
its minimal requirement for an effective field theory is stated as:

\medskip
\begin{screen}
Inside a given low-energy effective theory, 
there exists at least one realization satisfying the TCC condition \eqref{eq.TCC}.
\end{screen}
\medskip

Let us consider the possibility that we live in a metastable vacuum with a positive energy. 
Imposing the TCC condition \eqref{eq.TCC} on the trans-Planckian modes at the present time, 
it has been concluded in Ref. \cite{Bedroya:2019snp} that our Universe should terminate within a finite ``termination time" $t_{\rm end}$,
whose upper limit was given by $t_{\rm end}\lesssim  H_0^{-1}\log(M_{\rm Pl}/H_0)$.\footnote{This is the
cosmological counterpart of the scrambling time, see Ref.~\cite{Almheiri:2019yqk} for a recent related discussion.}
A stronger bound of the termination time, $t_{\rm end}\lesssim  H_0^{-1}\log(M_{\rm Pl}/H_{\rm inf})$, is obtained
by requiring the TCC condition \eqref{eq.TCC} on the trans-Planckian modes at the beginning of inflation.
\footnote{This condition is derived by combining $e^{N_{\rm inf} + N_{\rm after} + H_0 t_{\rm end}} \simeq M_{\rm Pl}/H_0$ with the horizon-crossing condition \eqref{eq:nlower}.
To be exact, the definition of $t_{\rm end}$ here is different from the former case. 
The termination time $t_{\rm end}$ is measured from the present time here 
but from the beginning of the current accelerated expansion for the former one. 
However, this difference is $\sim H_0^{-1}$ and negligible.}
Suppose that the present vacuum decays into another vacuum with a mean lifetime $\tau$.
From the argument above, one may argue that OTCC excludes the theories with metastable vacua whose mean lifetime $\tau$ is larger than the termination time $t_{\rm end}$. 

This conclusion, however, is not correct ---
the decay of a vacuum, quantum tunneling to another vacuum, occurs probabilistically and there is always a non-zero probability that a vacuum decays in a short time, $p_{\rm decay}(t_{\rm end}) = 1-e^{-t_{\rm end}/\tau}$. 
In formulating OTCC, we claimed that our Universe can be maximally atypical.
Therefore, as long as the probabilisitic interpretation of quantum mechanics is taken, 
it is possible to evade the upper limit on the termination time $t_{\rm end}$ 
simply by assuming that we are living in such a short-lived realization. 
This means that OTCC only requires a finite mean lifetime $\tau<\infty$ and there is essentially no constraint on it. 
As far as $\tau$ is finite, it is possible for the metastable vacuum to decay before the TCC violation.
Interestingly, the Standard Model is completely compatible with OTCC, as the electroweak vacuum is unstable.
The estimated lifetime of the electroweak vacuum is extremely huge \cite{Degrassi:2012ry,Buttazzo:2013uya,Chigusa:2017dux,Chigusa:2018uuj}, but has a finite value.

A similar argument opens up the possibility of the old inflation scenario \cite{Kazanas:1980tx, Guth:1980zm, Sato:1980yn}.
The old inflation scenario is usually excluded by the graceful-exit problem. 
However, there is a tiny probability that our patch of the Universe simultaneously decays into the Big Band phase after a sufficiently long period of inflation. 
Therefore, once one accepts our atypicality, the old inflation scenario is not excluded yet. 

These discussions illustrate an unsatisfactory point in the formulation of OTCC. 
At first sight, OTCC seems to give a proper swampland condition --- 
it selects the theories that allow at least one TCC-consistent realization. 
However, 
in order not for us to observe TCC violations, 
we should additionally assume that our Universe is one of such TCC-consistent realizations.
It is not clear apriori {\it why} this should be satisfied in the absence of any selection rule of realizations.
If it is impossible to justify this extreme atypicality of us, 
only possibility is to allow a small probability to observe the TCC violation,
at the cost of making the trans-Planckian censorship more loose.
In the next section, we will consider such a formulation of TCC, the probabilistic interpretation of TCC.  

\section{Probabilistic TCC} \label{s:ptcc}

Instead of complete eliminations, 
let us here allow for a small fraction of TCC-violating realizations and assign probabilities to observe them. 
One approach to formulate TCC along these lines is to appeal to
our usual intuition that our examples of TCC-violating realizations are probabilistically rare. 
This motivates us to propose the following probabilistic interpretation of TCC:

\waku{Probabilistic TCC (PTCC)}{the TCC condition \eqref{eq.TCC} is obeyed in almost all realizations, except that TCC-violating realizations with ``small" probabilities are tolerated.}

\begin{figure}[htbp]
    \centering
   \includegraphics{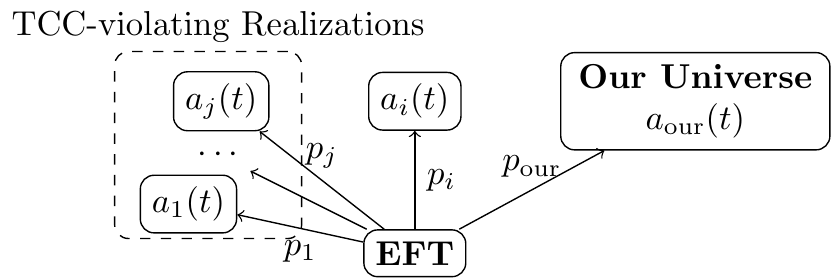}
    \caption{An example of EFT in the swampland for PTCC. PTCC claims that the TCC condition \eqref{eq.TCC} is obeyed in almost all realizations except that TCC violating realizations with ``small" probabilities are tolerated. In this figure, PTCC is violated when the probabilities $p_1, \dots, p_j$ for TCC-violating possibilities (given by scale factors $a_1(t),\dots, a_j(t)$) are not sufficiently ``small'' under a given probability measure. This inevitably introduces the dependence on the probability measure.}
    \label{fig:PTCC}
\end{figure} 

This version of TCC removes the previous objection: 
it does not exclude the Standard Model of particle physics 
because the examples of TCC-violating realizations discussed above have very small probabilities. 
This version could in principle work as a swampland condition, a restriction of effective field theories.
It is, however, unsatisfactory as a trans-Planckian ``censorship'' constraint, since the ``censorship" is incomplete.
From the phenomenological standpoint, 
PTCC can effectively work as TCC with a selection rule of realizations but no violation of principles of quantum mechanics.
For theories allowed in PTCC, 
cutting off TCC-violating realizations barely changes predictions for average values of observable, 
such as the primordial power spectra, 
because rare realizations do not contribute much to the average.

A problem of this proposal is that it depends on how to define the ``small'' probabilities ---
a larger threshold of probability eliminates a larger number of effective field theories. 
How rare a realization is accepted, and why? 

The threshold is also directly relevant to some observable predictions. 
For example, as already mentioned, 
TCC claims that 
a metastable de Sitter vacuum must terminate by a termination time $t_{\rm end} < T \simeq H_0^{-1} \ln (M_{\rm Pl}/H_{\rm inf})$. 
For the present value of the Hubble parameter $H_0$, the upper limit $T$ is much larger than the observed duration of the current accelerated expansion.
Let us consider the case that the present metastable vacuum will decay into another vacuum with a mean lifetime $\tau$.
In this case, the probability of survival at $t=T$, or the TCC violation, is estimated to be $p_{\rm survival}(T) = e^{-T/\tau} \simeq (H_{\rm inf}/M_{\rm Pl})^{1/(H_0 \tau)}$.
Therefore, 
if we take the threshold larger than $H_{\rm inf}/M_{\rm Pl} = {\cal O}(10^{-5})\sqrt{r}$, 
the lifetime should be smaller than the observed duration of the current accelerated expansion $\tau \lesssim t_{\rm age} \sim H_0^{-1}$. 
Then, we may have already observed signs of the vacuum decay. 

More fundamentally, PTCC depends on the choice of the probability, 
inevitable leading to the measure problem, the non-uniqueness of probabilities in cosmological setups  (see e.g.\ Refs. \cite{Guth:2000ka,Vilenkin:2006xv,Winitzki:2006rn,Linde:2007fr} for reviews).
It is possible to come up with some measure by choosing a particular time slice (e.g.,\ by proper-time or conformal time),
however this breaks the covariance of the measure, and the different choices can lead to vastly different probabilities.

Even when the probability is defined, 
there is a separate question: 
{\it typicality} of our Universe. \footnote{Typicality (``principle of mediocracy'' in the language of Ref.~\cite{Vilenkin:1994ua}) in cosmological setups 
have been discussed in the literature. See e.g.\ Refs.~\cite{Aguirre:2004qb,Weinstein:2005ef,Hartle:2007zv,Page:2007bt,Page:2008mx} for a variety of viewpoints on typicality.} 
In the formulation of PTCC, 
it is implicitly assumed that our Universe is typical 
--- if the probability of the TCC-violating realization is small, 
we rarely observe it. 

Is it appropriate to simply assume that our Universe is typical? 
Let us discuss an example for possible atypicality of our Universe. 
First, it can be argued that PTCC disfavors the dynamical relaxation of the cosmological constant 
through quantum tunnelings in the landscape of metastable de Sitter vacua, 
because it will experience a large number of de Sitter phases and eventually violate the TCC condition \eqref{eq.TCC}.
If so, 
we have to assume that our Universe has a tiny value of the cosmological constant from the initial time, 
which would be extremely ``unlikely" in the string landscape. 
This means that if our typicality is strictly assumed, 
we are excluding our own Universe 
---we concluded that we should not be present.

The typicality highly depends on the choice of the probability measure. \footnote{Our typicality is not guaranteed under some probability measures, 
as pointed out in Refs.~\cite{Guth:2000ka,Vilenkin:2006xv,Winitzki:2006rn,Linde:2007fr}.}
One possible way out is to introduce anthropic factors to the probability measure. 
Suppose that in formulating PTCC we use the probability measure
which is highly peaked when ``we'' are present. 
This anthropic modification could make our presence typical. 

However, it introduces another ambiguity in the formulation of PTCC. 
It is far from clear what is meant by ``we''---is this an intelligent life,
and then what counts as ``intelligence life''? 
We can tentatively propose some criteria, 
however, 
it is not clear {\it why} we should adopt such a measure. 
The problem is especially sharp when we remember that we wish to formulate PTCC as a swampland conjecture. If anthropic considerations are needed for formulating a swampland conjecture,
it seems we are arguing that our existence is strictly required for the consistency of UV completion in quantum gravity.
This is certainly a radical idea about quantum gravity.

\section{Is TCC a Swampland Conjecture?} 

In this paper, we discussed the question of whether or not the Trans-Planckian Censorship Conjecture (TCC) qualifies as a swampland conjecture. 

We first emphasized that the concept of swampland universality, i.e.
that a swampland conjecture constrains {\it any} possible scenario in a given effective field theory,
as long as it has a non-zero probability of realization, 
irrespective of whether or not it is realized in our Universe. 
This means that TCC can be applied to Universes which we do not necessarily observe, 
and hence can lead to more stringent constraints than are hitherto discussed in the literature.

We have seen that TCC formulated with the swampland universality, Universal TCC (UTCC), leads to contradictions
with the well-established theory, the Standard Model of particle physics and cosmology. 
A possible way out is to introduce a selection rule that eliminates all TCC-violating realizations.
For example, we can restrict to classical configurations, as formulated in Classical TCC (CTCC).
However, this leads to contradictions again with the Standard Model of particle physics. 
It would be difficult to find a successful selection rule, keeping the law of quantum mechanics.

We also discussed the possibility of applying TCC to our particular Universe, Observational TCC (OTCC). 
It is a swampland conjecture that picks up theories with at least one TCC-consistent Universe.
At first sight, it would give the TCC constraints in the literature. 
However, we have argued that this is not the case for the decay of a metastable de Sitter vacuum, 
or more generally when probabilities inherent in quantum mechanics are involved with the cosmological evolution.
This is because we now allow for maximal atypicality of us --- we should be living in one of the TCC-consistent Universes in order not to observe the violation of the TCC condition \eqref{eq.TCC}.
In this formulation, there remains a problem how to justify this atypicality of us.

We have therefore introduced a probabilistic interpretation to TCC, Probabilistic TCC (PTCC).
In this proposal, a small probability of the TCC violation is tolerated. 
This version could still work as a swampland condition, which eliminates theories with large probabilities of the TCC-violating realizations, 
although it is unsatisfactory as a trans-Planckian censorship constraint.

Note that all these different versions of TCC lead to different conclusions for practical problems.
For example, 
concerning the presence and the lifetime of a metastable de Sitter vacuum, the consequences are:
\begin{description}
  \item[UTCC/CTCC] The metastable de Sitter vacuum is not allowed for any mean lifetime.
  \item[OTCC] Any finite value of the mean lifetime is allowed. 
  \item[PTCC] The mean lifetime has a finite upper limit depending on the definition of the probability.
  
\end{description}

As is clear from this example, PTCC depends non-trivially on the choice of the probability measure.
We pointed out that PTCC suffers from the measure problem, i.e.\ the non-uniqueness of the probability measure,
and that the measure would likely be anthropic. One might still hope that  unknown (or even known) knowledge of quantum gravity naturally and inevitably picks up a unique measure, hence solving the measure problem. If this is the case, then at least the concept of probability is well-defined.

Irrespective of the measure problem, once we accept the probabilistic interpretation,  
then it seems very likely that 
swampland conjectures {\it in general} should be formulated probabilistically under the same measure 
--- the violation of the conjectures is allowed with a small probability.
This is a rather dramatic consequence. 
For example, 
the swampland conjecture of no exact global symmetry \cite{Misner:1957mt,Banks:1988yz,Polchinski:2003bq,Banks:2010zn,Harlow:2018tng} 
is believed to be a statement applicable to any possible realization in the landscape. 
However, the probabilistic nature of the swampland conjecture, if generally applied, 
suggests that some theories can survive if they predict a small probability of symmetry non-conserving processes.

In any formulation, we have seen that there are several challenges in considering TCC as a swampland condition ---
one needs to introduce any one of (unknown) selection rule, our atypicality (i.e.\ anthropic selection rule), or the TCC violation with a small probability.
These do not immediately reject TCC but their justification will require deeper understanding of quantum gravity. 
One could also argue that TCC is not a swampland condition after all, 
i.e.\ a constraint not on theories but just solutions. 
While this could be possible in principle, 
it requires modification of possible predictions extracted from a given theory --- 
this is more radical requirement than the standard swampland conjectures. 

\section*{Acknowledgements} 
We would like to thank Matthew Johnson, Misao Sasaki, Neil Turok, Daisuke Yamauchi  and Beni Yoshida for related discussions. M.Y.\ would like to thank Perimeter institute for hospitality during the final stages of this project.
This research was supported in part by WPI Research Center Initiative, MEXT, Japan, and by the JSPS Grant-in-Aid for Scientific Research  (R.S: No.~17K14286, No.~19H01891, S.S: No.~17H02878, No.~18K13535, No.~19H04609, M.Y: No.~17KK0087,
No.~19K03820, No.~19H00689).

\bibliographystyle{apsrev4-1}
\bibliography{swampland_2019_10}
\end{document}